\documentclass[12pt]{iopart}

\usepackage{dcolumn}% Align table columns on decimal point
\usepackage{bm}% bold math
\usepackage{ifpdf}
\usepackage{hyperref}
\usepackage{dcolumn}
\usepackage{bm}
\usepackage[spanish,english]{babel}
\usepackage{amsfonts}
\usepackage{amssymb}
\usepackage{graphicx}
\usepackage[latin1]{inputenc}
\usepackage{xcolor}
\usepackage{cite}

\begin{document}

\title{Geometric inequivalence of metric and Palatini formulations of General Relativity}

\author{Cecilia Bejarano$^1$, Adria Delhom$^2$, Alejandro Jim\'{e}nez-Cano$^3$, Gonzalo J. Olmo$^{2,4}$, Diego Rubiera-Garcia$^5$}

\address{$^1$Instituto de Astronom\'ia y F\'isica del Espacio (IAFE, CONICET-UBA),
Casilla de Correo 67, Sucursal 28, 1428 Buenos Aires, Argentina}
\address{$^2$Departamento de F\'{i}sica Te\'{o}rica and IFIC, Centro Mixto Universidad de Valencia - CSIC.
Universidad de Valencia, Burjassot-46100, Valencia, Spain}
\address{$^3$Departamento de F\'{i}sica Te\'{o}rica y del Cosmos and Centro Andaluz de F\'{i}sica de Part\'{i}culas Elementales Facultad de Ciencias, Avda Fuentenueva s/n,
Universidad de Granada, 18071 Granada, Spain}
\address{$^4$Departamento de F\'isica, Universidade Federal da
Para\'\i ba, 58051-900 Jo\~ao Pessoa, Para\'\i ba, Brazil}
\address{$^5$Departamento de F\'{i}sica Te\'orica and IPARCOS, Universidad Complutense de Madrid, E-28040
Madrid, Spain}

\eads{\mailto{cbejarano@iafe.uba.ar}, \mailto{adria.delhom@uv.es}, \mailto{alejandrojc@ugr.es}, \mailto{gonzalo.olmo@uv.es}, \mailto{drubiera@ucm.es}}

\begin{abstract}
Projective invariance is a symmetry of the Palatini version of General Relativity which is not present in the metric formulation. The fact that  the Riemann tensor changes nontrivially under projective transformations implies that, unlike in the usual metric approach, in the Palatini formulation this tensor is subject to a gauge freedom, which allows some ambiguities even in its scalar contractions. In this sense, we show that for the Schwarzschild solution there exists a projective gauge in which the (affine) Kretschmann scalar, $K\equiv {R^\alpha}_{\beta\mu\nu}{R_\alpha}^{\beta\mu\nu}$, can be set to vanish everywhere. This puts forward that the divergence of curvature scalars may, in some cases, be avoided by a gauge transformation of the connection.
\end{abstract}

\section{Introduction}

It is a widely accepted fact that General Relativity (GR) can be seen as an effective theory which will be superseded by a possibly quantum version when the curvature reaches the Planck scale. This is well illustrated by the Schwarzschild space-time,
\begin{equation}\label{eq:SchSol}
ds^2=-\left(1-\frac{2M}{r}\right)dt^2+\frac{1}{\left(1-\frac{2M}{r}\right)}dr^2+r^2(d\theta^2 +  \sin^2 \theta \ d\phi^2) \ ,
\end{equation}
which is the spherically symmetric solution of the vacuum Einstein equations, $G_{\mu\nu}=0$ with total mass $M$. Some components of this line element diverges at $r=2M$ and at $r=0$. Unveiling the nature of these divergences was crucial to fully understand the physics of black holes. In this sense, since GR is a diffeomorphism invariant theory, one can choose new coordinates (for instance, Eddington-Finkelstein) such that the metric singularity at $r=2M$ turns out to be avoidable. The $r=0$ divergence, on the contrary, cannot be removed by coordinate transformations because the coordinate-independent Kretschmann scalar explodes there as $K\equiv {R^\alpha}_{\beta\mu\nu}{R_\alpha}^{\beta\mu\nu}={48M^2}/{r^6}$.
This argument is generally used to conclude that a Schwarzschild black hole has a genuine curvature singularity at $r=0$ \cite{allbooks1,allbooks2,allbooks3,allbooks4,allbooks5,
allbooks6,allbooks7,allbooks8,allbooks9,allbooks10}. The unbounded curvature suggests that classical GR should  break down when $K \sim 1/l_P^4$ (where $l_P \equiv \sqrt{\hbar G/c^3}$ is Planck's length), where the quantum gravitational degrees of freedom are expected to play a non-negligible role. The Schwarzschild solution, therefore, encapsulates the beginning and the end of GR, in the sense that it illustrates both the novelties of the theory and its physical limitations.

Given the importance of the physical limitations of the theory already raised by the Schwarzschild solution, it is worth considering if different realizations of GR are also affected by the same problems, since this could help envision new ways to tackle the quantum gravity issue. Among the various possibilities, the Palatini formulation can be regarded as the most natural one as, in fact, it has served as the starting point for several important developments such as the ADM formulation \cite{Arnowitt:1962hi}, supergravity \cite{allbooks2}, Deser's completion of a diffeomorphism invariant massless spin two theory from flat to curved space-time \cite{Deser:1969wk}, etc. Moreover, it is conceptually closer to the Einstein-Hilbert formulation than the teleparallel approaches (based on torsion or non-metricity with vanishing Riemann \cite{BeltranJimenez:2019tjy}), which facilitates the qualitative and quantitative comparison with respect to the standard framework.

The main aim of this work is thus to emphasize the geometric inequivalence between the metric and the Palatini formulations of GR and to put forward the relevance of the  symmetries associated to the connection in order to deal with curvature pathologies. In our opinion, this point has received little attention in the literature despite its potential implications for our understanding and interpretation of gravitational phenomena.

\section{Test particles, projective invariance, and the Palatini formulation of GR}

Rooted on the equivalence principle, whose experimental status still enjoys very good health (see e.g. \cite{Schlamminger:2007ht}), test particle paths in GR are determined by the geodesic equation, which in an arbitrary parametrization $\lambda$ takes the form
\begin{equation}\label{eq:geo_gen}
\frac{d^2 x^\mu}{d\lambda^2}+\Gamma^\mu_{\alpha\beta}\frac{d x^\alpha}{d\lambda}\frac{d x^\beta}{d\lambda}=f(\lambda)\frac{d x^\mu}{d\lambda} \ .
\end{equation}
where $\Gamma^\mu_{\alpha\beta}$ are the components of the connection. An affine parametrization $\tau(\lambda)$ is the case in which the right-hand side vanishes and, in general, it can be found by solving $f(\lambda)=\tau_{\lambda\lambda}/\tau_\lambda$, with $\tau_\lambda=d\tau/d\lambda$, which turns the above equation into
\begin{equation}\label{eq:geo_aff}
\frac{d^2 x^\mu}{d\tau^2}+\Gamma^\mu_{\alpha\beta}\frac{d x^\alpha}{d\tau}\frac{d x^\beta}{d\tau}=0 \ .
\end{equation}
In the affine parametrization we can say that the {\it fictitious force} term on the right-hand side of (\ref{eq:geo_gen}) has vanished, which provides optimal conditions for the analysis of test particles.  Obviously, since optimal experimental conditions are useful but not strictly necessary, physical paths are independent of the parametrization chosen in much the same way as curvature scalars are independent of the coordinates chosen. %Optimal experimental conditions are useful but not necessary . 

Though the affine parametrization may seem to split clocks in two classes, namely, those for which the right-hand side of (\ref{eq:geo_gen}) vanishes and those for which it does not, this is not quite so. In fact, when the paths in a given parametrization are known, it is always possible to find a one-form $\xi\equiv \xi_\alpha dx^\alpha$, whose components satisfy  $\xi_\alpha \frac{dx^\alpha}{d\lambda}=-f(\lambda)$, such that $\lambda$ becomes the affine parameter of a new connection
\begin{equation}\label{eq:projective}
\tilde{\Gamma}^\mu_{\alpha\beta}=\Gamma^\mu_{\alpha\beta} + \xi_{\alpha}\delta^{\mu}_{\beta} \ ,
\end{equation}
whose paths, obviously, coincide with those of the original $\Gamma^\mu_{\alpha\beta}$. In terms of $\tilde{\Gamma}^\mu_{\alpha\beta}$, Eq.(\ref{eq:geo_gen}) turns into
\begin{equation}\label{eq:geodesics_afflambda}
\frac{d^2 x^\mu}{d\lambda^2}+\tilde{\Gamma}^\mu_{\alpha\beta}\frac{d x^\alpha}{d\lambda}\frac{d x^\beta}{d\lambda}=0 \ .
\end{equation}
This shows that free particle paths are not uniquely associated to the connection $\Gamma^\mu_{\alpha\beta}$  but to a family of connections $\tilde{\Gamma}^\mu_{\alpha\beta}$ related to the original one via the so-called projective transformations (\ref{eq:projective})  \cite{Schouten,Eisenhart}. Indeed, given the form of the geodesic equation, this family of connections is simply another manifestation of the freedom we have in the  parametrization of a given path\footnote{It should be noted that the normalization of the tangent to a given path depends on the parametrization of the curve: $g_{\mu\nu} \frac{dx^\mu}{d\lambda}\frac{dx^\nu}{d\lambda}=s(d\tau/d\lambda)^2$, with $s=0,-1$ for null and time-like trajectories, respectively. For time-like paths, the choice of parametrization should be irrelevant as long as it is monotonical, i.e., $d\tau/d\lambda\neq 0$. The failure of this norm to be $-1$ can be seen as either an effect of the non-metricity ($Q_{\lambda\mu\nu} \equiv \nabla^\Gamma_\lambda g_{\mu\nu}\neq 0$) induced by the projective transformation or as due to the choice of a clock which does not measure the proper time.}.

Besides leaving invariant the paths followed by test particles,  projective transformations are also a symmetry of the Einstein-Palatini action
\begin{equation}\label{eq:action}
\mathcal{S}_{EP}=\frac{1}{16\pi G} \int d^4 x \sqrt{-g} g^{\mu\nu} R_{\mu\nu}(\Gamma)\ .
\end{equation}
To see this, we first note that the Riemann curvature tensor is mathematically defined in terms of a connection $\Gamma^\alpha_{\nu\mu}$, {\it a priori} independent of $g_{\mu\nu}$,  as ${R^\alpha}_{\mu\beta\nu}({\Gamma}) \equiv \partial_\beta \Gamma^\alpha_{\nu\mu}-\partial_\nu \Gamma^\alpha_{\beta\mu}+\Gamma^\alpha_{\beta\lambda}\Gamma^\lambda_{\nu\mu}-\Gamma^\alpha_{\nu\lambda}\Gamma^\lambda_{\beta\mu}$. From a field theory perspective,  ${R^\alpha}_{\mu\beta\nu}({\Gamma})$ can be seen as the field strength of $\Gamma^\alpha_{\nu\mu}$, and under the projective transformations (\ref{eq:projective}) it changes as
\begin{equation}\label{eq:RiemannTrans}
{R^\alpha}_{\mu\beta\nu}(\tilde{\Gamma})= {R^\alpha}_{\mu\beta\nu}(\Gamma)+\delta_{\mu}^{\alpha}F_{\beta\nu} \ ,
\end{equation}
where $F_{\mu\nu} \equiv \partial_{\mu}\xi_{\nu}- \partial_{\nu}\xi_{\mu}$ is the field strength of $\xi_\mu$. Since (\ref{eq:action}) only depends on the contraction of the metric with the Ricci tensor, $R_{\mu\nu}(\tilde{\Gamma})\equiv  {R^\alpha}_{\mu\alpha\nu}(\tilde{\Gamma})= {R}_{\mu\nu}(\Gamma)+F_{\mu\nu}$,
the antisymmetry of $F_{\mu\nu}$ guarantees the invariance of (\ref{eq:action}) under projective transformations: $g^{\mu\nu}R_{\mu\nu}(\tilde{\Gamma})=g^{\mu\nu}R_{\mu\nu}({\Gamma})$. This shows the unphysical (gauge) character of $\xi_\mu$, which neither affects the space-time metric equations/solutions nor test particle trajectories. Note, in this sense, that only the symmetric part of the Ricci tensor enters in the Einstein tensor, which implies that $G_{\mu\nu}(\tilde\Gamma)=G_{\mu\nu}(\Gamma)$.

We now recall that in the metric (or Einstein-Hilbert) formulation of GR, the metric is the only geometric field. The Palatini formulation, on the contrary, considers metric and connection as  equally fundamental and {\it a priori} logically independent geometric entities, in such a way that the form of the connection follows from solving the connection equation upon independent variations of the metric and the connection. The introduction of the affine connection as a fundamental field allows to enhance the symmetries of the theory from diffeomorphism symmetry alone to diffeomorphism plus projective symmetries. As a result, the most general solution for the connection is the Levi-Civita one plus an arbitrary projective mode. This indicates that the metric formulation of GR can be seen as a particular projective gauge fixing of the Palatini one \cite{Bernal:2016lhq,Janssen:2019htx}. Though these two formulations satisfy the same Einstein equations\footnote{Strictly speaking this is only true in vacuum or when minimally coupled to bosonic matter fields. For fermions, the connection picks up a projectively-invariant torsional term, but  that contribution will be irrelevant for the purposes of this work, as it is mainly focused on the vacuum Schwarzschild solution.}, the different symmetry properties that they possess turns out to be very relevant for their underlying geometric interpretation, as we will see next.

\section{Projective transformations and curvature invariants}

From  the transformation law (\ref{eq:RiemannTrans}), one readily sees that the affine Kretschmann scalar, $K\equiv {R^\alpha}_{\beta\mu\nu}{R_\alpha}^{\beta\mu\nu}$, is not invariant under projective transformations. In fact, it changes as
\begin{equation}
K (\tilde{\Gamma})= K(\Gamma) + 4F_{\mu\nu}F^{\mu\nu} \ .
\end{equation}
In light of this, one can consider the Schwarzschild solution (\ref{eq:SchSol}), for which $K(\Gamma)=\frac{48M^2}{r^6}$, %is given by (\ref{eq:Kret})
and take, for instance, $\xi_{\mu}=(\phi(r),0,0,0)$ with $\phi(r)=\pm \sqrt{3/2}M/r^2$ to find a physically equivalent description, with the same metric and the same geodesic paths, but in which $K (\tilde{\Gamma})$ identically vanishes everywhere.
From a physical perspective, given the symmetries of the action (\ref{eq:action}), the choice of projective gauge should be as irrelevant as the specific set of coordinates chosen to write the line element because the action and the field equations of the theory are invariant under arbitrary transformations of both of them. There is nothing in the Palatini version of the theory that singles out the purely metric gauge $\xi_\mu=0$ over the others, as there is no preferred choice of coordinates to write the line element.  Therefore, in the Schwarzschild solution of the Palatini formulation of GR, the curvature singularity of $K$ at $r=0$   is an artifact that can be avoided by changing the projective gauge from $\xi_\mu=0$ to  $\xi_{\mu}=(\pm \sqrt{3/2}M/r^2,0,0,0)$.

Related to this, note that curing the Kretschmann by shifting from  $\Gamma^\mu_{\alpha\beta}$ to $\tilde{\Gamma}^\mu_{\alpha\beta}$ induces divergences in the antisymmetric part of the corresponding Ricci tensor which, as shown above, lies in an unobservable (pure gauge) sector of the theory. The same applies to the associated torsion and non-metricity tensors induced by the projective transformation.

One should note, however, that though the affine Kretschmann scalar, ${R^\alpha}_{\beta\mu\nu}{R_\alpha}^{\beta\mu\nu}$, is not really a scalar under projective transformations, there is another quadratic invariant of the affine Riemann tensor which is indeed a (diffeomorphism and projective) scalar, namely, $P\equiv{R^{\alpha\beta}}_{\mu\nu}{R^{\mu\nu}}_{\alpha\beta}$. In the Schwarzschild solution, this quantity becomes $P=48M^2/r^6$ and diverges as $r\to 0$. The fact that the quadratic scalars $K$ and $P$ are degenerate in the metric formulation of GR but are radically different in the metric-affine formulation, puts into question the physical meaning of curvature invariants as proxies for pathological behaviors. In this sense, it is also worth noting that in nonvacuum space-times other types of projective-invariant curvature divergences associated to the Ricci and (symmetric) Ricci-squared scalars may arise. The insensitivity of such quantities to the projective modes may be related in this case to the fact that minimally-coupled matter fields are projectively invariant as well. This puts forward the different nature of such divergences (as compared to those of the Riemann tensor and the scalar $K$), which claims for a different approach for their regularization and/or interpretation. In the case of electrically charged black holes, for instance, the (symmetric) Ricci-squared scalar $R_{\mu\nu}R^{\mu\nu}$ diverges as $\sim q^4/r^8$ if Maxwell electrodynamics is assumed. This divergence is inherited from the matter stress-energy tensor, which diverges as $\sim q^2/r^4$ and, therefore, has infinite total energy. If nonlinear matter effects are considered, such as in Born-Infeld  electrodynamics \cite{Gibbons:1997xz}, the stress-energy divergence is weaker, $\sim 1/r^2$, which regularizes the total energy of the field but, nonetheless, generates a curvature divergence $R_{\mu\nu}R^{\mu\nu}\sim 1/r^4$. Whether or not this curvature divergence should be interpreted as a pathological geometric effect, despite describing a matter sector with bounded total energy, is a question that should be further explored \cite{Horowitz:1995ta}.

\section{Summary and discussion}

In this work we have used the classical Schwarzschild solution to illustrate that the metric and the Palatini versions of GR are not geometrically equivalent.  While in the metric formulation the Riemann tensor can be regarded as an observable associated to the (unique) Levi-Civita connection of the metric, in the Palatini version its physical status is more subtle because it is not invariant under all the symmetries of the theory. In fact, the existence of an extra symmetry associated to the connection implies that the affine Riemann curvature tensor in the Palatini version is subject to an additional gauge freedom which is not present in the standard metric approach. Although the existence of this symmetry was already well known in the literature \cite{Hehl:1981, Julia:1998, Julia:2000, Hall:2007, Dadhich:2012}, to our knowledge its impact on curvature invariants had not been studied yet. The Palatini approach has received significant interest in the last decade in relation with extensions of Einstein's theory of the $f(R)$ or Born-Infeld types (see \cite{Olmo:2011uz} and \cite{BeltranJimenez:2017doy} for comprehensive reviews), which are particular cases of the so-called Ricci-based gravities \cite{BeltranJimenez:2017doy,Afonso:2017bxr,Afonso:2018bpv,Afonso:2018hyj,Delhom:2019zrb,BeltranJimenez:2019acz,BeltranJimenez:2020ProjLargo}. The relation between Palatini and metric formulations of $f(R)$ has also been analyzed in many works  \cite{Capo2007,Capo2010,Capozziello:2010ih,Capozziello:2010ef,Dadhich:2012}.

 Exploiting the projective freedom, we have shown that the problems in the Schwarzschild metric (\ref{eq:SchSol}) at $r=2M$ are as empty of physical significance as the blowup of the affine Kretschmann scalar $K$ at $r=0$, since both can be removed by a suitable gauge choice of coordinates or of projective mode, respectively. The projective freedom, however, does not help in removing neither metric divergences nor curvature divergences of projective-invariant quantities,  such as that in the scalar $P\equiv{R^{\alpha\beta}}_{\mu\nu}{R^{\mu\nu}}_{\alpha\beta}$ in the Schwarzschild solution and others in the Ricci tensor in non-vacuum space-times.  Indeed, notice that since the metric version of GR can be seen as a (projective) gauge fixed Palatini GR, all quantities invariant under diffeomorphisms and projective transformations in the latter will be in one-to-one correspondence with the corresponding invariants of the former. Nonetheless, the possibility of gauging away some geometric infinities by considering alternative representations of the classical theory with an enhanced symmetry group is an encouraging new result that motivates the exploration of alternative realizations of Einstein's theory with extra symmetries encoded in fields other than the metric. This suggests a novel research avenue that will be further considered in the future. In particular, the potential construction of nontrivial covariant derivatives (other than the purely metric) leading to diffeomorphism and projective invariant quantities should be further explored.

To conclude let us point out that, since projective transformations neither change the form of the metric nor geodesic paths, they cannot cure the singular character of the Schwarzschild solution, which is geodesically incomplete. Geodesic completeness is essential to avoid the destruction of information or its creation out of nowhere (by naked singularities) and also for the very existence of observers, who are essential to perform measurements. Our findings about the possibility of removing some  curvature divergences by means of symmetry transformations further reinforces the idea that the blowup of curvature scalars should not be seen as the {\it reason} for the geodesic incompleteness on which the singularity theorems are based \cite{Penrose1,Hawking,Carter,Senovilla1,Senovilla2,Curiel} (see \cite{Geroch:1968ut,Earman:1995fv} for illuminating and in-depth discussions of the notion of a singular space-time). This has been recently verified in some explicit counter-examples in the Palatini formulation of modified theories of gravity \cite{Bejarano:2017fgz,Menchon:2017qed}, in which curvature divergences do not prevent geodesic completeness. In light of this, any argument based on the blowup of affine curvature scalars to estimate the scale at which quantum gravity effects are relevant (see our introduction above) might be misleading if the underlying space-time structure is not strictly Riemannian. A deeper understanding of this issue could help envisage new strategies in the search of an improved theory of matter and gravity free of the pathologies of the metric formulation of classical GR.

\section*{Acknowledgements}

C. B. is funded by the National Scientific and Technical Research Council (CONICET).
AD and AJC are supported by a PhD contract of the program FPU 2015 (Spanish Ministry of Economy and Competitiveness) with references FPU15/05406 and FPU15/02864, respectively.
GJO is funded by the Ramon y Cajal contract RYC-2013-13019 (Spain).
DRG is funded by the \emph{Atracci\'on de Talento Investigador} programme of the Comunidad de Madrid No. 2018-T1/TIC-10431, and acknowledges support from the Funda\c{c}\~ao para a Ci\^encia e
a Tecnologia (FCT, Portugal) research grants Nos.  PTDC/FIS-OUT/29048/2017 and PTDC/FIS-PAR/31938/2017.
This work is supported by the Spanish projects FIS2017-84440-C2-1-P, FIS2014-57387-C3-1-P (MINECO/FEDER, EU) and i-LINK1215 (CSIC), the project H2020-MSCA-RISE-2017 Grant FunFiCO-777740, the project SEJI/2017/042 (Generalitat Valenciana), the Consolider Program CPANPHY-1205388, and the Severo Ochoa grant SEV-2014-0398 (Spain).  This article is based upon work from COST Action CA15117 and CA18108, supported by COST (European Cooperation in Science and Technology).

\end{document}